\documentclass[superscriptaddress, aps, pre, reprint, floatfix]{revtex4-2}
\usepackage{graphicx,hyperref,color,upgreek}
\usepackage{amsmath}
\usepackage{latexsym}
\usepackage{float}
\usepackage{amssymb}
\usepackage{multirow}
\usepackage{graphicx}
\usepackage{textcomp}
\usepackage{hyperref}
\usepackage{array}
\usepackage{xcolor}

\begin{document}
\title{Growth and aging in a few phase-separating active matter systems}
\author{Florian Dittrich}
\affiliation{Institute of Physics, Johannes Gutenberg-Universit\"{a}t, Mainz, Germany}
\author{Jiarul Midya}
\affiliation{Theoretical Physics of Living Matter, Forschungszentrum J\"{u}lich, 52425 J\"{u}lich, Germany}
\author{Peter Virnau}
\email{virnau@uni-mainz.de}
\affiliation{Institute of Physics, Johannes Gutenberg-Universit\"{a}t, Mainz, Germany}
\author{Subir K. Das}
\email{das@jncasr.ac.in}
\affiliation{Theoretical Sciences Unit and School of Advanced Materials, Jawaharlal Nehru Centre for Advanced Scientific Research, Jakkur P.O., Bangalore 560064, India}
\date{\today}
%
\begin{abstract}
Via computer simulations we study evolution dynamics in systems of continuously moving Active Brownian Particles. The obtained results are discussed against those from the passive $2D$ Ising case. Following sudden quenches of uniform configurations to state points lying within the miscibility gaps and to the critical points, we investigate the far-from-steady-state dynamics by calculating quantities associated with structure and characteristic length scales. We also study aging for quenches into the miscibility gap and provide a quantitative picture for the scaling behavior of the two-time order-parameter correlation function. The overall structure and dynamics are consistent with expectations from the Ising model. This remains true for certain active lattice models as well for which we present results for quenches to the critical points.
\end{abstract}
\maketitle

\section{Introduction}
\label{sec:intro}
Nonequilibrium models of self-propelled or active particles describe a multitude of phenomena ranging from the movement of bacteria and artifical microswimmers to macroscopic flocks of birds ~\cite{Vicsek:2012,Bechinger:2016,roadmap}. Some of these systems exhibit cooperative phenomena such as motility-induced phase separation (MIPS)~\cite{Cates:2015} that resembles the passive liquid-gas phase separation but occurs in absence of any attractive interactions: At large propulsion speeds and low rotational diffusion, artificial microswimmers can self-trap and form clusters. The resulting phase diagram shows a binodal curve of coexisting densities that ends in a critical point. Computationally, artificial microswimmers are often studied with continuously moving active Brownian particles (ABPs)~\cite{Fily:2012,Redner:2013,Stenhammar:2013,Stenhammar:2014,Wysocki:2014,Bialke:2015,Siebert:2017,Digregorio:2018} or variants thereof, but recently active lattice models have gained attention as well ~\cite{Whitelam:2018,Partridge:2019,Dittrich:2021}.

Whether the phase behavior in the vicinity of a non-equilibrium critical point is unique, and if it belongs to any standard universality class is a question of fundamental interest and has sparked an ongoing controversy ~\cite{Siebert:2018,Caballero:2018,Partridge:2019,Maggio:2020,Dittrich:2021}: For ABPs, a determination of the critical point and its associated critical exponents revealed results incompatible with any known universality class~\cite{Siebert:2018}, while active Ornstein-Uhlenbeck particles appear to be compatible with $2D$-Ising behaviour \cite{Maggio:2020}. Similarly, active lattice models exhibit exponents close to the $2d$-Ising values~\cite{Partridge:2019} even though small model-dependant deviations remain~\cite{Dittrich:2021}. For the description of static critical behavior first theoretical approaches have appeared recently which may reconcile some of these discrepancies \cite{Speck:2022}.

In this manuscript we focus on dynamical aspects of active systems. Understanding of non-equilibrium dynamics following quenches of homogeneous systems to the critical point, as well as to state points inside the coexistence region, is of fundamental as well as practical relevance \cite{Bray:2002, Puri:2009}. In the context of passive matter systems associated phenomena received much attention. In this broad area, recent focus has been on active matter systems \cite{cre:2014, das2:2017, KBinderSoftMaterials2021, Gregoire:2004, landau_binder_2005, chak:2020, paul:2021}. In a class of studies the objective is to understand the scaling behaviors related to structure, growth and aging \cite{Bray:2002, DFisher:1988, Liu:1991, Desai:1996, lifshitz:1961, Jiarul:2015, das2:2017}. Below we provide brief descriptions of these non-equilibrium aspects.

Typically, growth in such non-equilibrium situations, following quenches inside the coexistence region, occurs in a power-law fashion, viz., average size of domains, rich or poor in particles of a particular type, $\ell$, grows with time ($t$) as \cite{Bray:2002}
\begin{equation}
\ell \sim t^{\alpha}.
\label{eq:glaw}
\end{equation}
Such a growth is usually self-similar in nature, i.e., the domain patterns at two different times  are different from each other only via a change in $\ell$. This property is reflected in the scaling behavior of the two-point equal time correlation function \cite{Bray:2002},
\begin{equation}
C(r,t)=\left<\psi(\vec{r},t)\psi(0,t)\right> - \left<\psi(\vec{r},t)\right>\left<\psi(0,t)\right>,
\label{eq:cofr}
\end{equation}
as \cite{Bray:2002}
\begin{equation}
C(r,t)\equiv C(r/\ell).
\label{eq:scl_cofr}
\end{equation}
Here $\psi$ is a space ($\vec{r}$) and time dependent order parameter. Another important property associated with such non-equilibrium systems is the aging phenomena. This can be captured in the relaxation behavior of the two-time order-parameter auto-correlation function \cite{DFisher:1988}
\begin{equation}
C_{\rm ag}(r,t)=\left<\psi(\vec{r},t)\psi(\vec{r},t_w)\right> - \left<\psi(\vec{r},t)\right>\left<\psi(\vec{r},t_w)\right>,
\label{eq:auto_cr}
\end{equation}
where $t_w$ ($< t$) is a waiting time, also referred to as the age of the system. As opposed to the equilibrium systems, the time translation invariance in growing systems is not obeyed. Thus, $C_{\rm ag}$ does not exhibit collapse of data from different values of $t_{\rm w}$ when plotted versus $t-t_w$, but is reported to scale
as a function of $t/t_{\rm w}$ as \cite{DFisher:1988}
\begin{equation}
C_{\rm ag} \sim \left(\frac{t}{t_w}\right)^{-\alpha\lambda},
\label{eq:cag_scl}
\end{equation}
$\lambda$ being referred to as an aging exponent.

Similar interest exists for quenches to the critical point. In this case the correlation in the system is expected to grow with time as \cite{Hohenberg:1977, Onuki:2002, landau_binder_2005}
\begin{equation}
\xi(t) \sim t^{1/z},
\label{eq:xi_vs_t}
\end{equation}
$z$ being a dynamic critical exponent. Note that in the long time limit values of $\xi$ diverge with the approach to the critical point in a power-law fashion with an exponent $\nu$ \cite{Fisher_1967}. 

Obtaining the values of $\alpha$, $\lambda$ and $z$ are of fundamental importance in the domain of dynamics of phase transitions. Understanding of these are quite advanced for various lattice systems in the case of passive matter. For fluids, the status is reasonably poor. In the case of active matter systems, such interest is very recent. In this work, we intend to obtain these quantities for phase separating systems consisting of Active Brownian particles \cite{Siebert:2018, Cates:2015}. In addition, we also study active lattice systems \cite{Partridge:2019, Whitelam:2018} as they are computationally less demanding and thus yield statistically better data. These systems did show interesting deviations in steady-state critical behaviour from ABPs in prior work \cite{Dittrich:2021}. Therefore, a comparative analysis of dynamical behaviour adds further understanding towards the uniqueness of active matter systems.

Note that self-propelled particles forming non-equilibrium active systems offer a wide range of interesting behaviour and applications \cite{Vicsek:2012,Bechinger:2016,roadmap,Das:2014,Trefz:2016}. While phase transition and the overall nonequilibrium behaviour in these systems constitute a broad field of ongoing research \cite{Fily:2012,Redner:2013,Stenhammar:2013,Stenhammar:2014,Wysocki:2014,Bialke:2015,Siebert:2017,Digregorio:2018,Siebert:2018}, we add an additional aspect of non-equilibrium behavior by quenching uncorrelated homogeneous active systems to correlated or phase separated states.

\begin{figure}[h]
\includegraphics[width=0.9\linewidth]{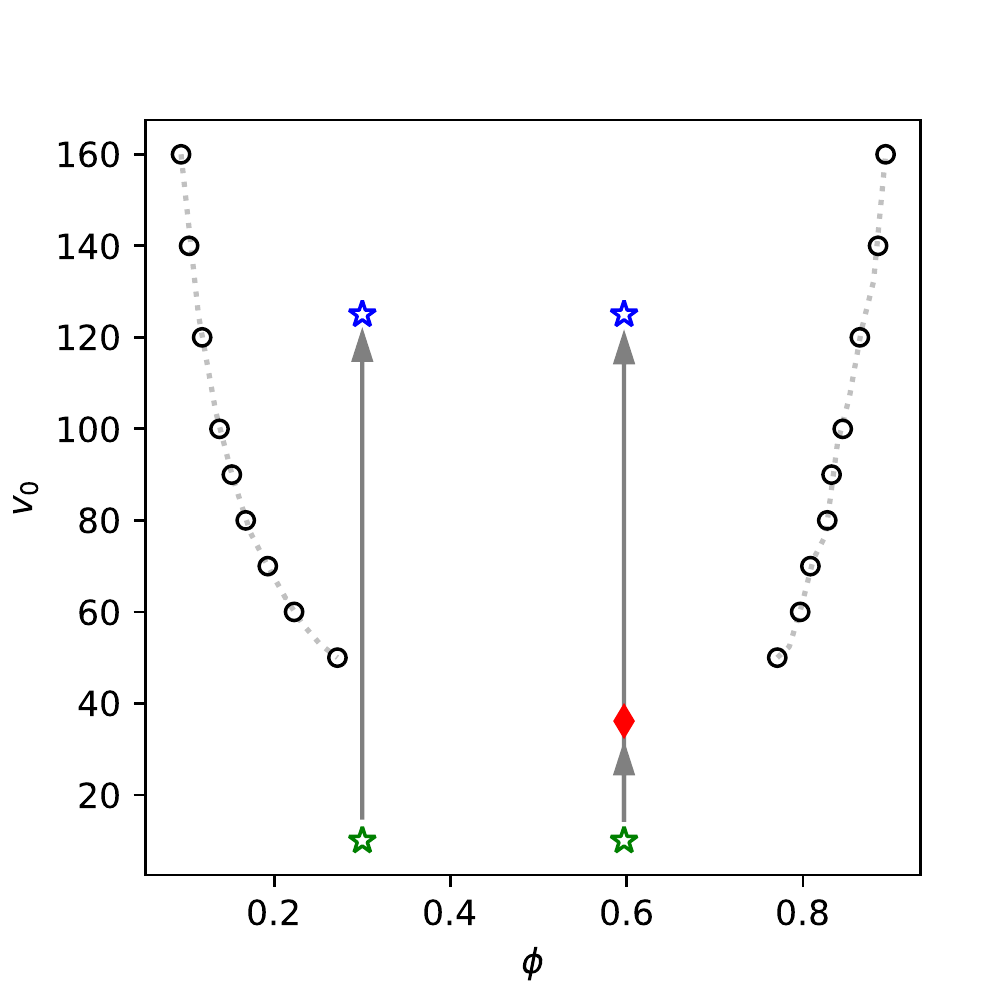}
\caption{Phase diagram for ABPs as shown in Ref.~\cite{Siebert:2018}. The green stars mark the initialisation points from where the systems were quenched at constant packing fractions $\phi=0.597$ and $\phi=0.3$ into the phase separated region (blue stars) and to the critical point (red diamond). 
Dotted lines were only drawn to guide the eye.}
\label{fig:phdia} 
\end{figure}

\section{Methods}
\subsection{Model and simulations: Active Brownian Particles}
\label{sec:modelABPs}
Systems of Active Brownian Particles in two dimensions consist of hard disks which are actively propelled along their orientation (see below). Periodic boundary conditions are applied in both dimensions and equations of motion are given by \cite{Siebert:2018}:
\begin{equation}
 \dot{\mathbf r}_k = -\frac{D_t}{k_BT}\nabla_k U + v_0\left(
\begin{array}{c}
\cos\varphi_k \\ 
\sin\varphi_k
\end{array}\right) + \sqrt{2D_t}\mathbf R_k,
\label{eq:dyn}
\end{equation}
where $\mathbf R_k$ is normal distributed Gaussian noise, $D_t$ the translational diffusion constant, and $U$ arises from a purely repulsive Weeks-Chandler-Andersen (WCA) potential between disks with $\epsilon = 100$ and $\sigma = 1$ as in Ref.~\cite{Siebert:2018}. 
If not mentioned otherwise, units are from now on omitted and correspond to standard simulation units. Furthermore $D_t$ is set to 1. With a cutoff distance of $r = 2^{1/6}$ we obtain an effective hard disk Barker-Henderson diameter $d_\text{BH} \approx 1.10688$.  A particle's orientation is described by the angle $\varphi_k$, which undergoes free rotational diffusion with diffusion coefficient $D_r$, i.e. $\dot{\varphi_k}=\sqrt{2D_r}R_r$, where $R_r$ is Gaussian distributed, has unit variance and is neither correlated between particles nor in time. Each particle is propelled along its orientation with constant speed $v_0$.
GPU-based simulations were performed using HOOMD-blue \cite{Anderson:2020} applying a Brownian integrator with a time step of $10^{-6}$. Temperature was set to 1, and simulations were performed at fixed volumes and particle numbers. If the rotational diffusion $D_r$ (set to $3D_t/d_{\mathrm{BH}}^2 \approx 2.45$ throughout this work) is small with respect to the active velocity $v_0$, a self-trapping mechanism can be observed \cite{Buttinoni:2013, Cates:2015}: Particles that form an emerging cluster require more time to orient away from the cluster than it takes for other particles to reach and enlarge it. This leads to a separation into a dense and a dilute phase and a non-equilibrium phase diagram with a critical point (Fig.~\ref{fig:phdia}) \cite{Siebert:2018} even in the absence of explicit attractions.

In the present work we have performed several quenches into the phase-separated region and to the critical point. All simulations started in a mixed state with $v_0 = 10$ and were first equilibrated for $2\times10^7$ time steps (corresponding to 20 MD times). For the quenches into the phase-separated region the final active velocity was set to $v_0 = 125$. Critical density ($\phi=0.597$ \cite{Siebert:2018}) was established in a system of size $1024\times1024$ with 649636 particles (Fig.~\ref{fig:SABP}b), while the quench to the low density branch (at $\phi=0.3$) (Fig.~\ref{fig:SABP}a) was realized with 314573 particles. To improve statistics we averaged over 10 independent runs each. 

For quenches to the critical point (Fig.~\ref{fig:SABP}c) the final active velocity was set to $v_0 = 40/d_\text{BH} \approx 36.14$ \cite{Siebert:2018}. In order to study the scaling of the steady-state correlation length $\xi_{\rm max}$ with system size, different system sizes had to be realized. In particular, 100 independent runs for size $64\times64$ containing 2500 particles each, 50 runs for size $128\times128$ with 10000 particles, and 15 runs for size $256\times256$ containing 40401 particles each were performed.

\subsection{Model and simulations: Active lattice systems}

Quenches to the critical points were also performed for three active lattice models which are described in detail in Ref.~\cite{Dittrich:2021}. In contrast to ABPs \cite{Siebert:2018}, these systems are already reported to exhibit steady state critical exponents close to the $2D$-Ising values \cite{Dittrich:2021}. From a computational point of view they are also less demanding, and superior statistics can be achieved in a straightforward implementation on CPUs.

Rotational diffusion and active propulsion are handled similarly to ABPs, but parameters are probabilities or rates in Monte Carlo (MC) simulations. Each particle can occupy a single site and is oriented towards one of its neighboring sites. Density is defined as the number of occupied divided by the total number of sites. Again, all simulations were performed with a fixed number of particles in 2D with periodic boundary conditions. One Monte Carlo time unit consists of as many individual Monte Carlo attempts as there are lattice spaces in the system. A rotation move only changes the orientation of the particle. In a translation move, a particle attempts to move to a neighboring site, which is always accepted if the targeted space is empty and rejected otherwise. In order to implement activity, movements along the particles' orientation were chosen with higher probability. Other directions were also allowed with reduced probability to account for translational diffusion.

As in Ref. \cite{Dittrich:2021} Model I \cite{Partridge:2019} on a hexagonal (hex.) and a square (sq.) lattice were investigated to study the influence of lattice geometry on emerging dynamical properties. In each simulation step, the program attempts to change the orientation of a particle first: A Gaussian distributed random number (having standard deviation $\sigma_{\rm rot}$ and zero mean) is chosen and rounded to the nearest integer. The current orientation is adjusted by that integer, and the move is accepted with a probability 1. As indicated, the rotational diffusion parameter $\sigma_{\rm rot}$ governs the width of the Gaussian distribution and hence the activity of the model: A low value for $\sigma_{\rm rot}$ corresponds to a low probability for orientation adjustments and thus enhanced activity and vice versa. Afterwards, a translation move is attempted with the same particle and accepted if the destination location is empty. The direction for the move is chosen at random, with probability $w_+$ along the particle's current orientation and with probability $w_t$ for any of the remaining directions. For the hexagonal lattice probabilities are set to $w_+=25/30$ and $w_t=1/30$ \cite{Partridge:2019}, for the square lattice $w_+=17/20$ and $w_t=1/20$ \cite{Dittrich:2021}. 

In Model II \cite{Whitelam:2018} either a rotation or a translation move is attempted in an individual simulation step on the square lattice. A clockwise or anticlockwise rotation is performed with rate $w_1=0.1$, an attempted move along the current orientation is undertaken with rate $w_+$ and in any other direction with $w_t=1$. Activity is regulated by adjusting $w_+$. Probabilities for each move are obtained by dividing the individual rates by the sum of all rates, namely ($w_+ + 3 w_t + 2 w_1$). For a more in-depth discussion of the lattice models including steady state critical exponents and visualizations of particle moves, the reader is referred to Ref.~\cite{Dittrich:2021}.

All lattice systems were equilibrated at the corresponding critical densities ($0.524$ for Model I hex., $0.498$ for Model I sq. and $0.527$ for Model II \cite{Dittrich:2021}) for 5,000 time units in a mixed state and then quenched to the critical points. For equilibration, activity in Model I was set to $\sigma_{rot}=1$ and in Model II to $w_+ = 1.25$. $200$ independent runs were performed for $L=512$. The system size for the hexagonal lattice was increased by $2/\sqrt{3}$ to 592 in one direction to account for the hexagonal structure. 
Quenches to the critical points were simulated for five different system sizes. 400 independent runs were undertaken for $L=64$ and $L=128, 200$ runs for $L=256$ and $512$ and $50$ for $L=1024$. For the hexagonal lattices, one dimension was again adjusted as described above. Critical simulation parameters were taken from Ref.~\cite{Dittrich:2021} as $\sigma_\text{rot}=0.3048$ for Model I hex., $\sigma_\text{rot}=0.2415$ for Model I sq. and $w_\text{+}=4.76$ for Model II sq.

\section{Results and discussion}
\label{sec:Results}
As already stated, in Fig.~\ref{fig:phdia} we show the phase diagram of the off-lattice model. The MIPS phase behaviour resembles that of a vapor-liquid phase separation in passive systems rather closely. Nevertheless, it is not clear yet whether the critical behaviour can be attributed to the Ising universality class. On the one hand, a study with the model used in this paper showed clear deviations from $2D$ Ising universality class \cite{Siebert:2018}. On the other hand, studies using other models concluded agreement with $2D$ Ising universality class \cite{Partridge:2019,Maggi:2020}. This led to some controversy, as generic arguments of renormalization~\cite{Hohenberg:1977} suggest that all these models should fall into the same universality class. However there are other considerations which suggest that nonequilibrium models might fall somewhere in between because of an unstable fixed point of the renormalisation group (RG) flow \cite{Caballero:2018,Speck:2022}.
%
%
\begin{figure}[ht!]
\includegraphics[width=1.0\linewidth]{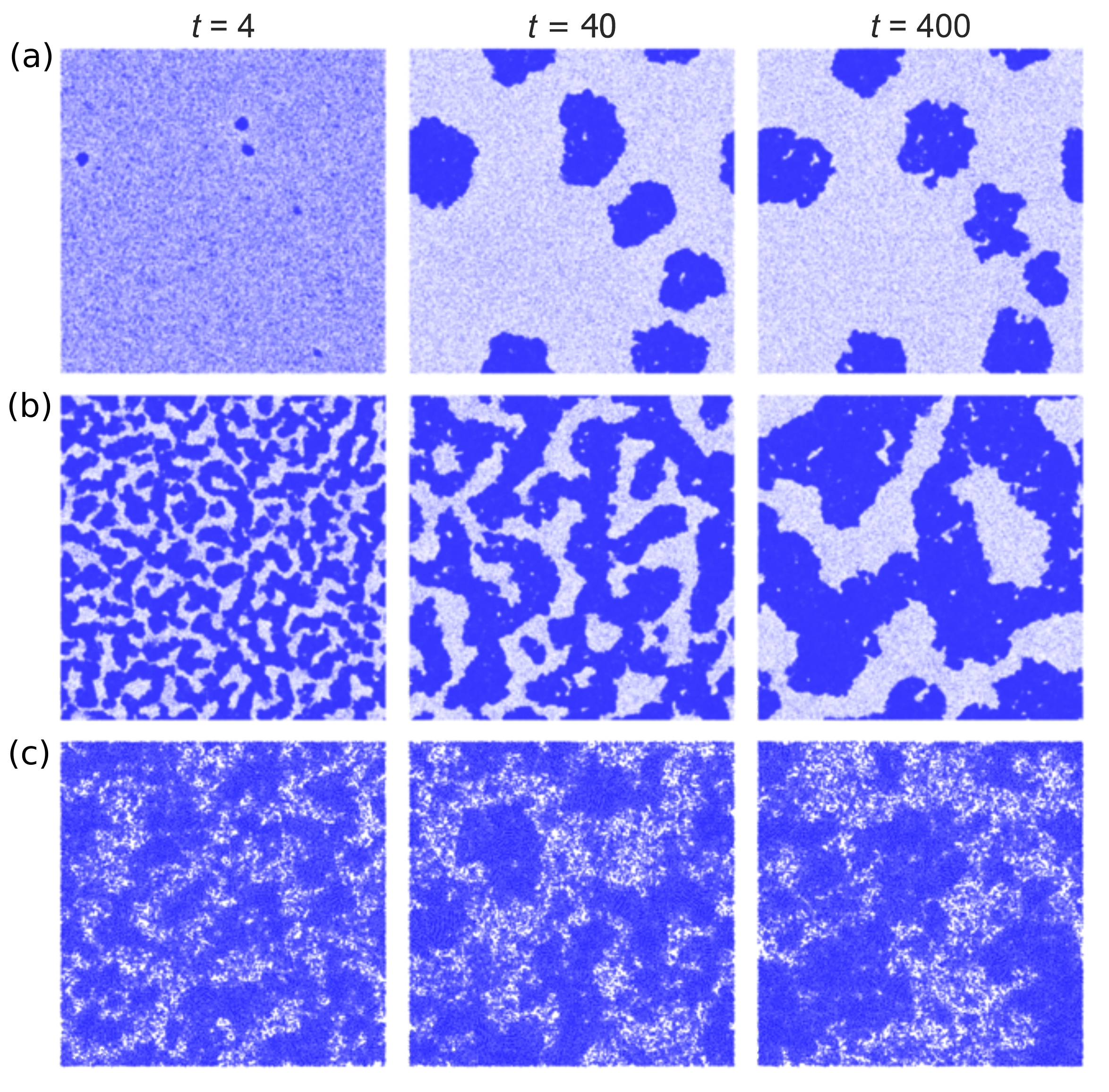}
\caption{Snapshots obtained during the evolutions of the ABP model, at three different times, after the quenches took place. Time $t$ is given in MD units. (a) For density $\phi = 0.3$, we have simulated system size $1024\times1024$ following quench to $v_0 = 125$. (b) For critical density $\phi = 0.597$, we have system size $1024\times1024$ and quench was to $v_0 = 125$. (c) For critical density $\phi = 0.597$, system size was $256\times256$ and quench was to the critical $v_0 = 40/d_\text{BH} \approx 36.14$.}
\label{fig:SABP} 
\end{figure}
%
\begin{figure}[ht!]
\includegraphics[width=0.8\linewidth]{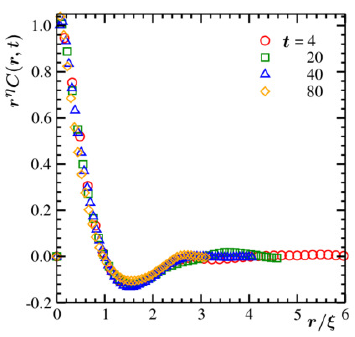}
\caption{Scaling plots of the correlation functions for the quench protocol of Fig.~\ref{fig:SABP}(c). Data from several different times are included for quenches of random initial configurations to the critical point, for the off-lattice model.}
\label{fig:corrfn_offLattice_cr}
\end{figure} 
\begin{figure}[ht!]
\includegraphics[width=0.8\linewidth]{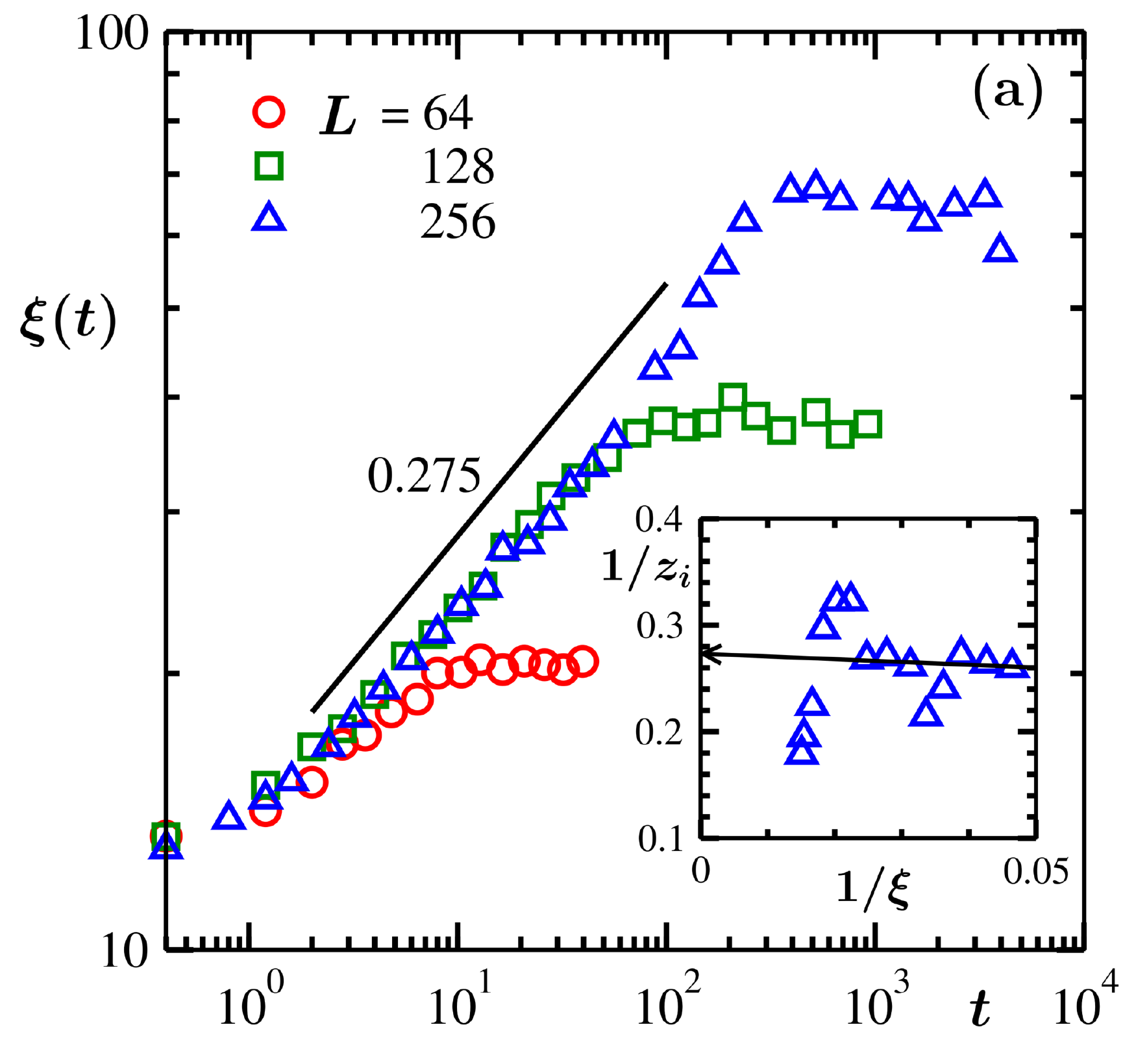}
\includegraphics[width=0.8\linewidth]{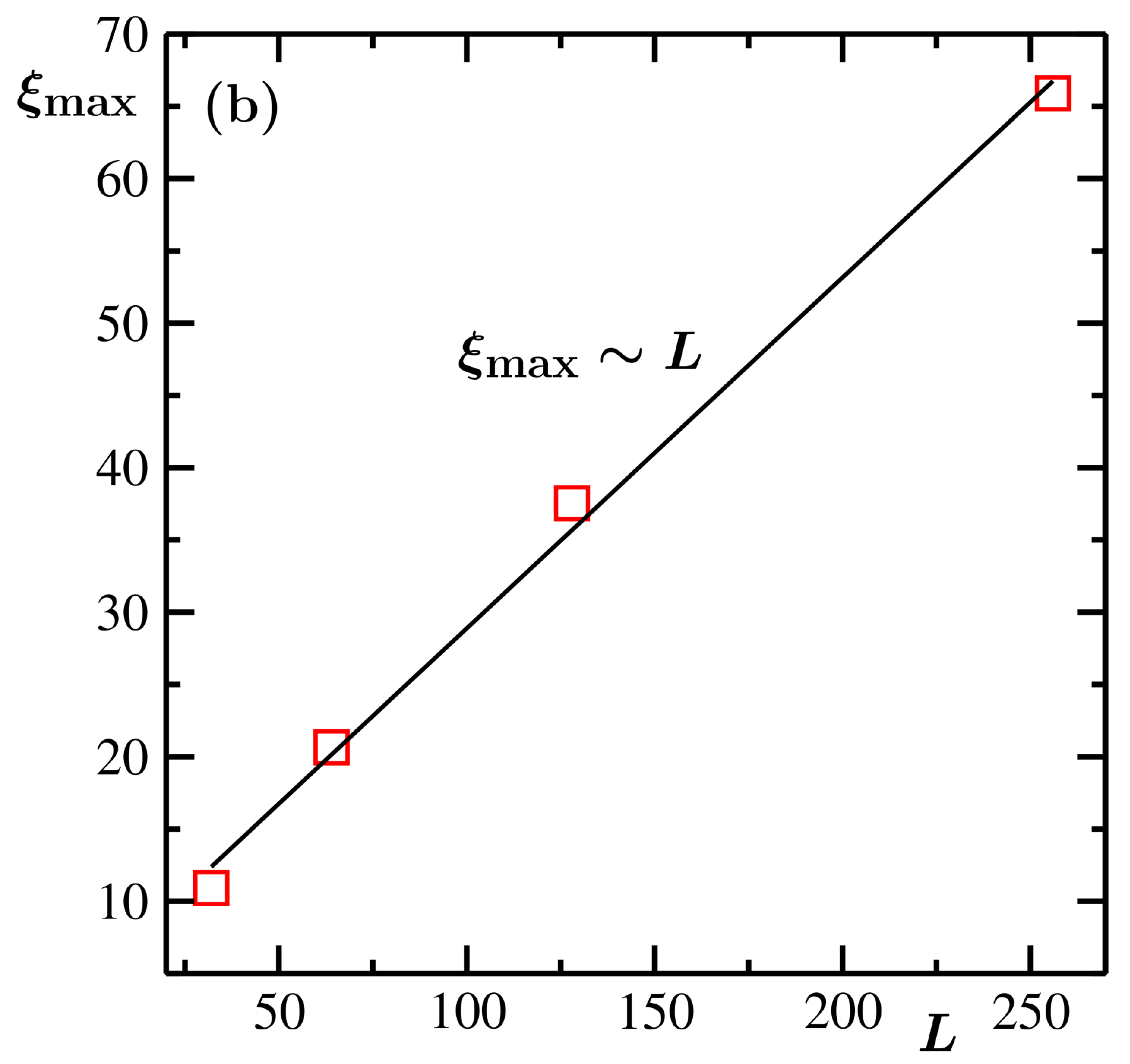}
\caption{(a) Results for growing correlation length are shown from different system sizes for quenches of random initial configurations to the critical point. The solid line is a power-law, the value of the exponent mentioned next to it. Inset shows the plot of instantaneous exponent, $1/z_i$, as a function of $1/\xi$, for $L=256$. (b) The steady state values of the correlation length, $\xi_{\rm max}$, are plotted versus the system size $L$. These results are for the continuum model.}
\label{fig:dLen+xiMax-offLatt-cr} 
\end{figure}
\begin{figure}[ht!]
\includegraphics[width=0.8\linewidth]{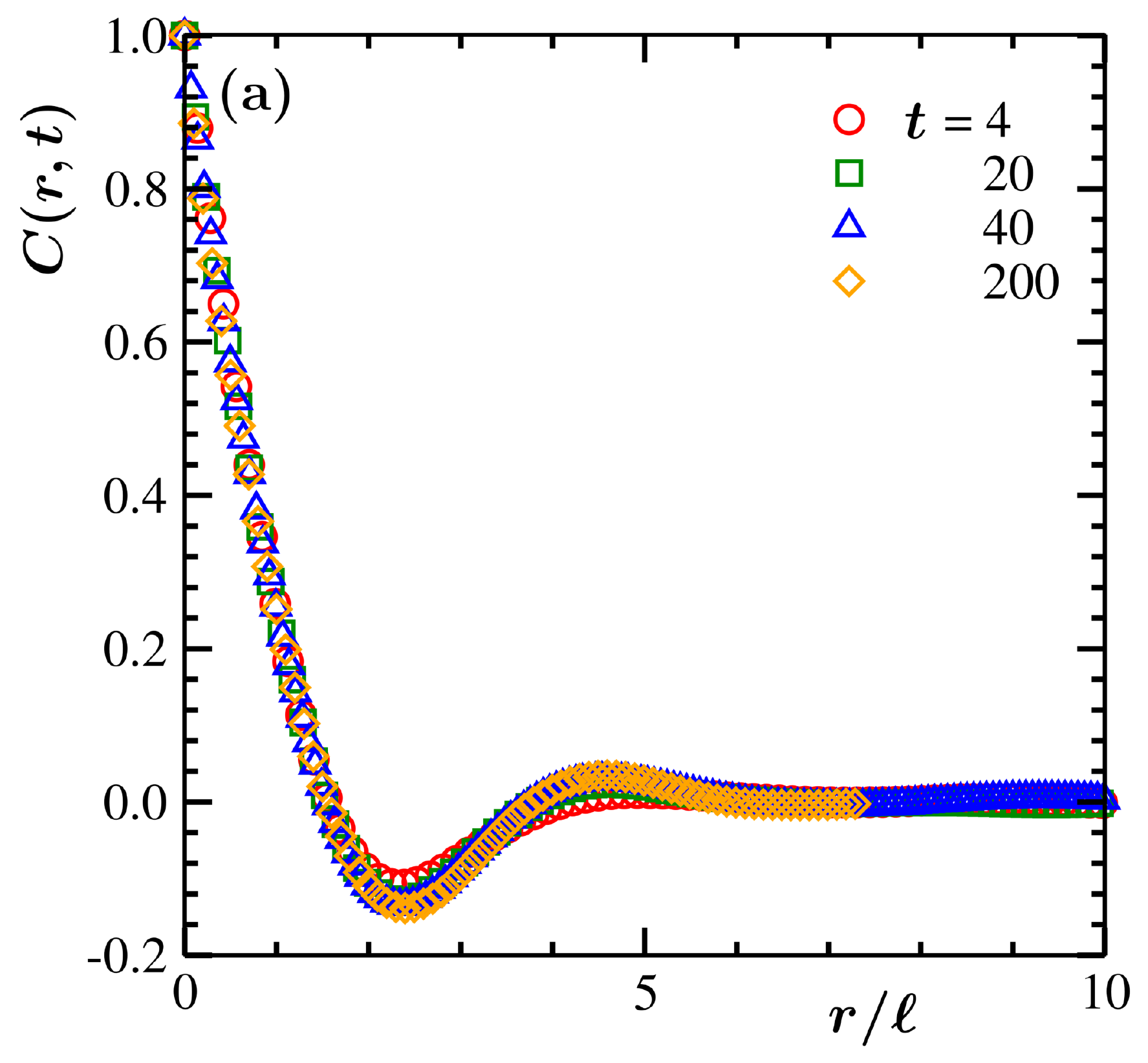}
\includegraphics[width=0.8\linewidth]{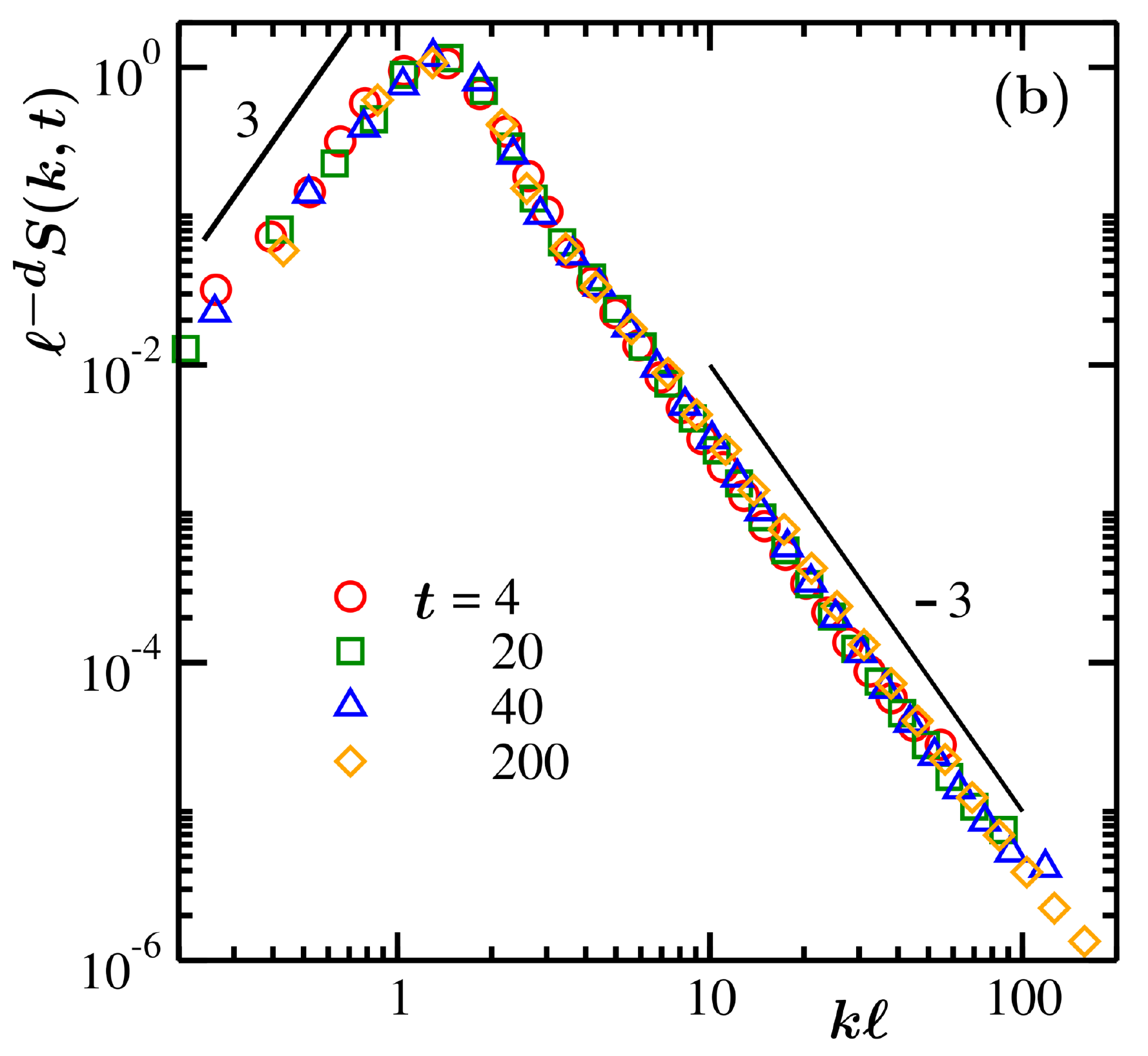}
\caption{Scaling plots of (a) $C(r,t)$ and (b) $S(k,t)$, for the quenching protocol described in Fig.~\ref{fig:SABP}(b). In part (b) the solid lines represent  power-laws with exponent values mentioned in adjacent locations. These results are from simulations of the off-lattice model.}
\label{fig:scaled-corrfn+sstfrac-offLatt-cr}
\end{figure}
\begin{figure}[ht!]
\includegraphics[width=0.8\linewidth]{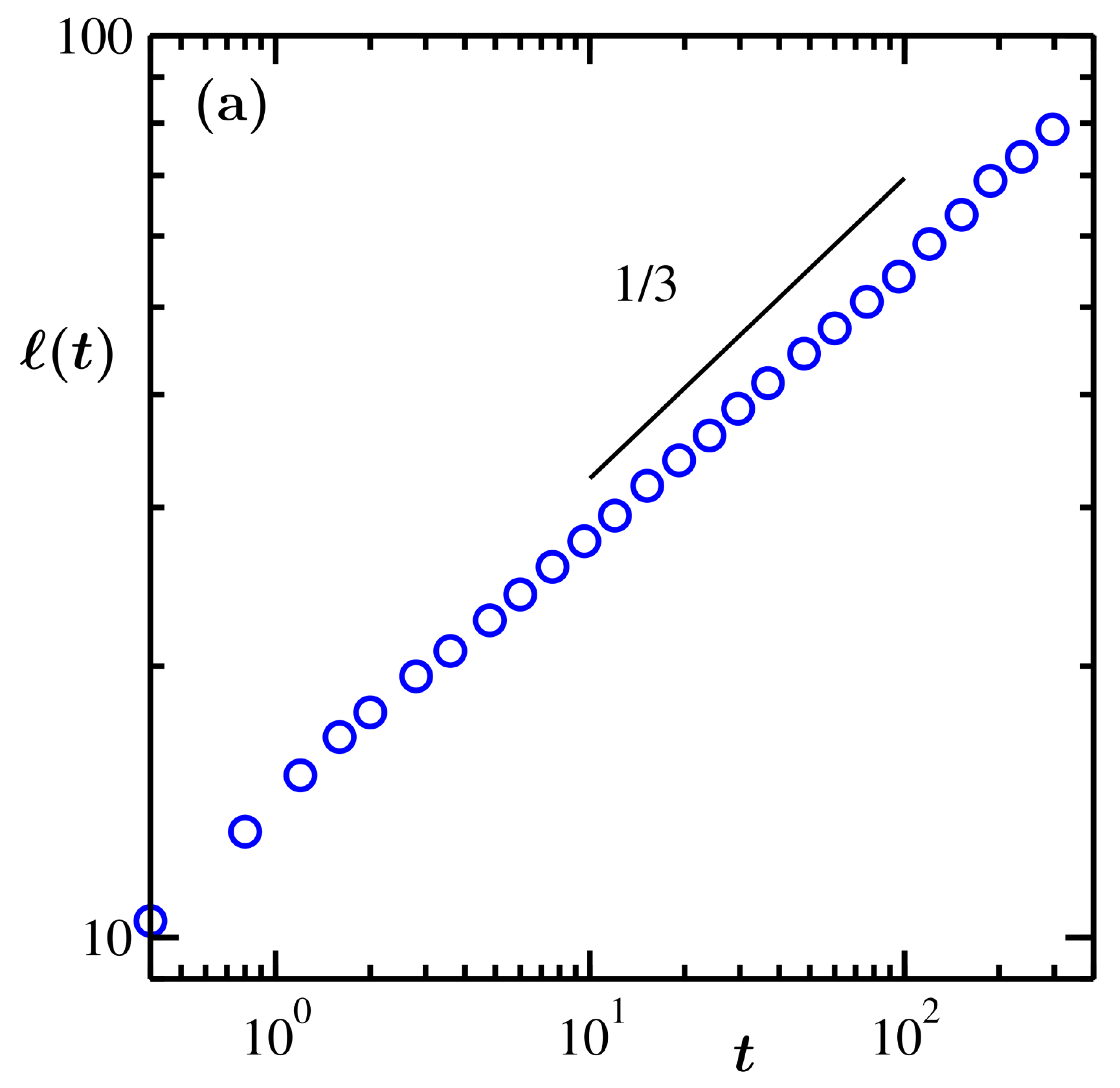}
\includegraphics[width=0.8\linewidth]{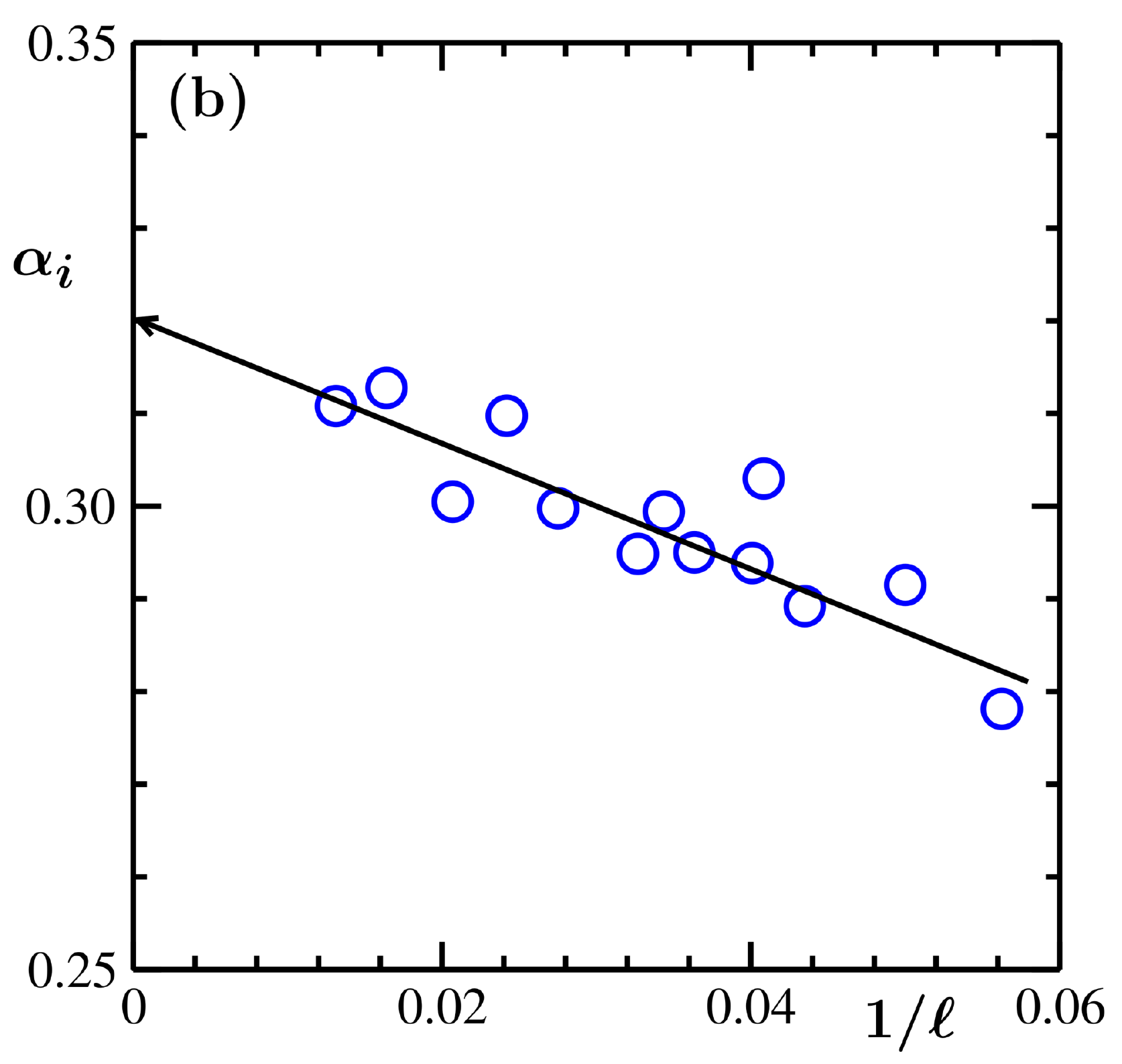}
\caption{(a) Plot of the average domain length as a function of time for the quenching protocol described in Fig.~\ref{fig:SABP}(b). (b) Instantaneous exponent corresponding to the growth in (a) is plotted versus $1/\ell$. These results are from the simulations of the continuum model.}
\label{fig:dLeng+insExp-offLatt-ps}
\end{figure}
\begin{figure}[ht!]
\includegraphics[width=0.8\linewidth]{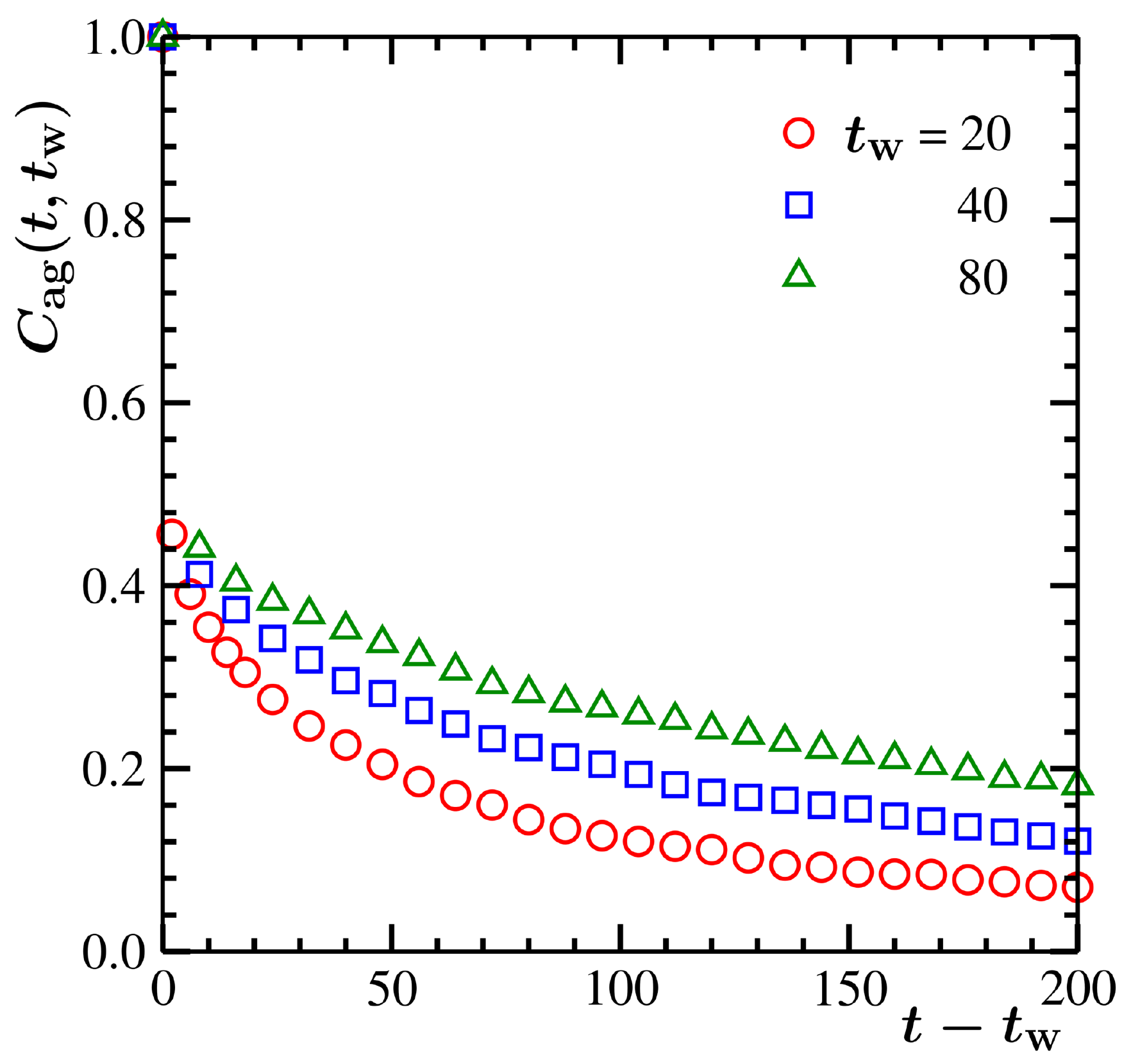}
\caption{Autocorrelation functions from the simulations of the off-lattice model are plotted versus the translated time $t-t_{\rm w}$, for a few values of the waiting time $t_{\rm w}$. These results are for the protocol of Fig.~\ref{fig:SABP}(b).}
\label{fig:autocorrFn-offLatt-ps}
\end{figure}
\begin{figure}[ht!]
\includegraphics[width=0.8\linewidth]{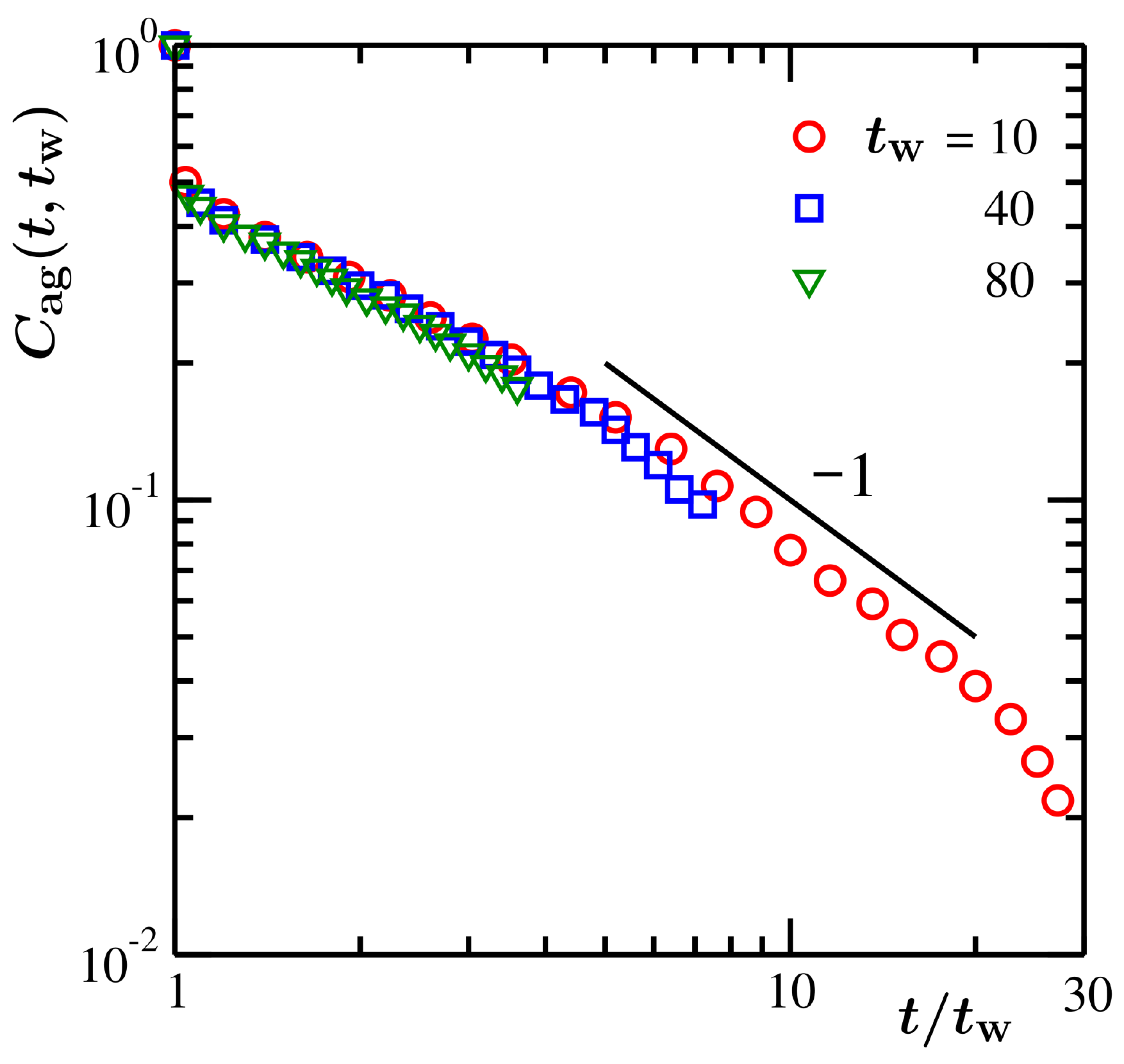}
\caption{Same as Fig.~\ref{fig:autocorrFn-offLatt-ps} but here we show $C_{\rm ag}(t,t_{\rm w})$ as a function of $t/t_{\rm w}$. The solid line has a power-law decay. The value of the exponent is mentioned next to the line.}
\label{fig:scaled-autocorrFn-offLatt-ps}
\end{figure}
\begin{figure}[ht!]
\includegraphics[width=0.8\linewidth]{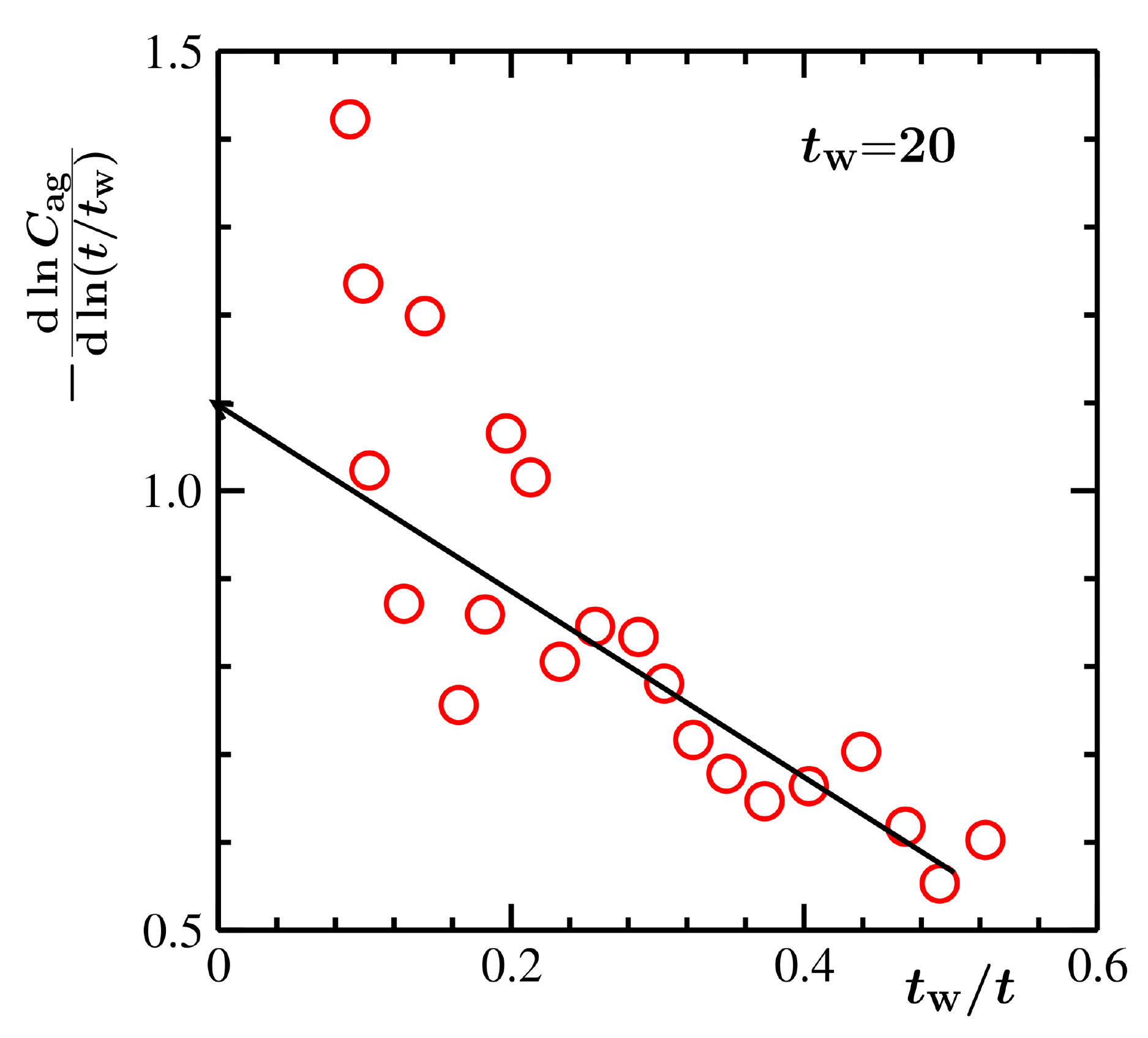}
\caption{The instantaneous aging exponent is plotted versus $t_{\rm w}/t$ for a data set in Fig.~\ref{fig:scaled-autocorrFn-offLatt-ps}. The solid line is a guide to the eyes.}
\label{fig:insExp-autocorrFn-offLatt-ps}
\end{figure}
\begin{figure}[ht!]
\includegraphics[width=1.0\linewidth]{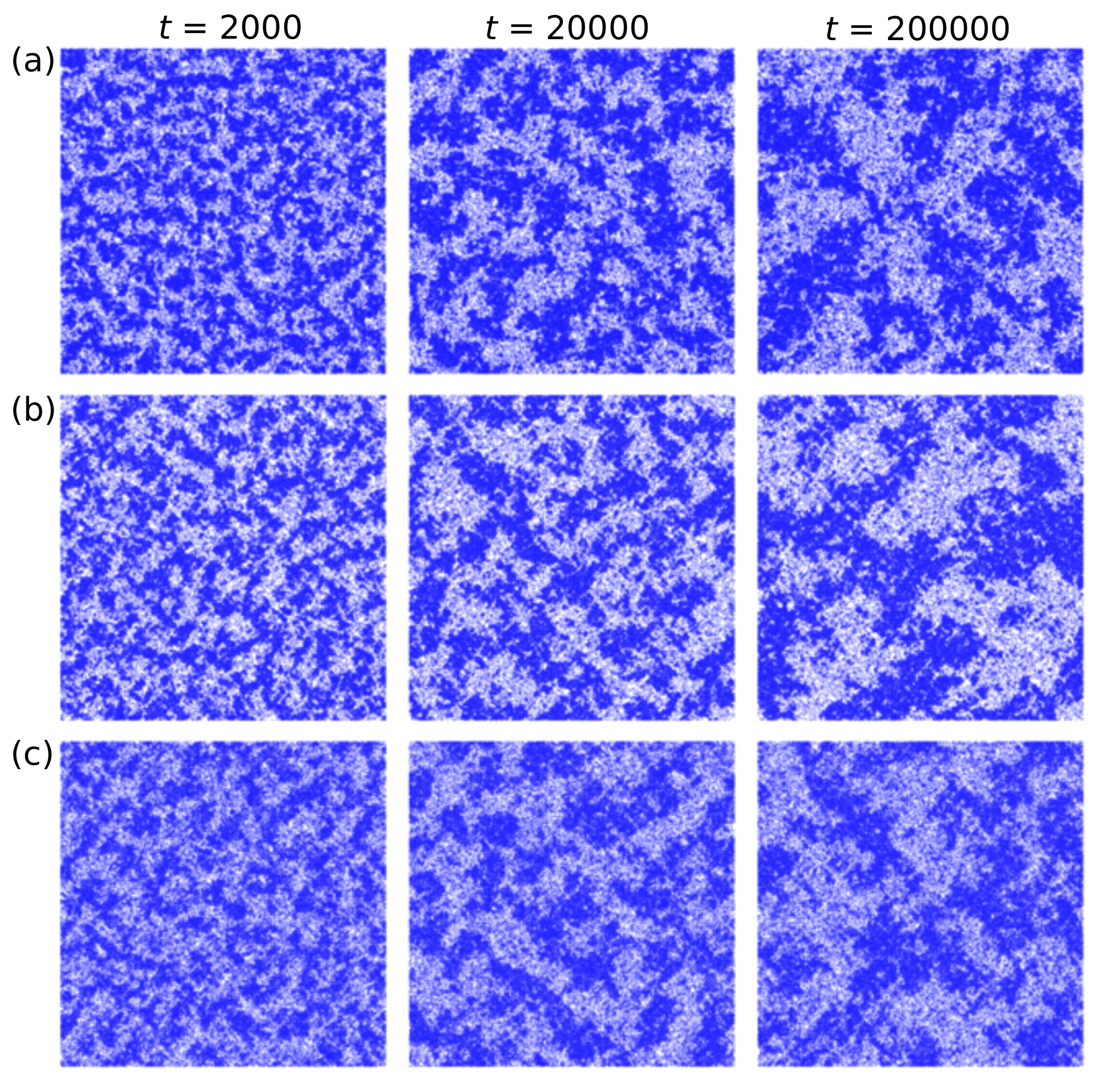}
\caption{Snapshots obtained during the evolutions of the considered lattice models are presented for three different times after the quenches to the critical points took place. Time $t$ is given in MC steps. The locations of the particles are marked. (a) For Model I hex., a system at critical density 0.524, is quenched to $\sigma_\text{rot}=0.3048$. The system is of size $512\times592$, to adjust for the hexagonal lattice structure. (b) For Model I sq., a critical density (0.498) system is quenched to $\sigma_\text{rot}=0.2415$. The system is of size $512\times512$. (c) For Model II sq., a system with critical density 0.527 is quenched to $w_\text{+}=4.76$. System is of size $512 \times 512$. The comparison of the results with the 2D conserved Ising model is presented in Fig.~S4, in the Supplementary Material.}
\label{fig:SLCR}
\end{figure}
\begin{figure}[ht!]
\includegraphics[width=0.8\linewidth]{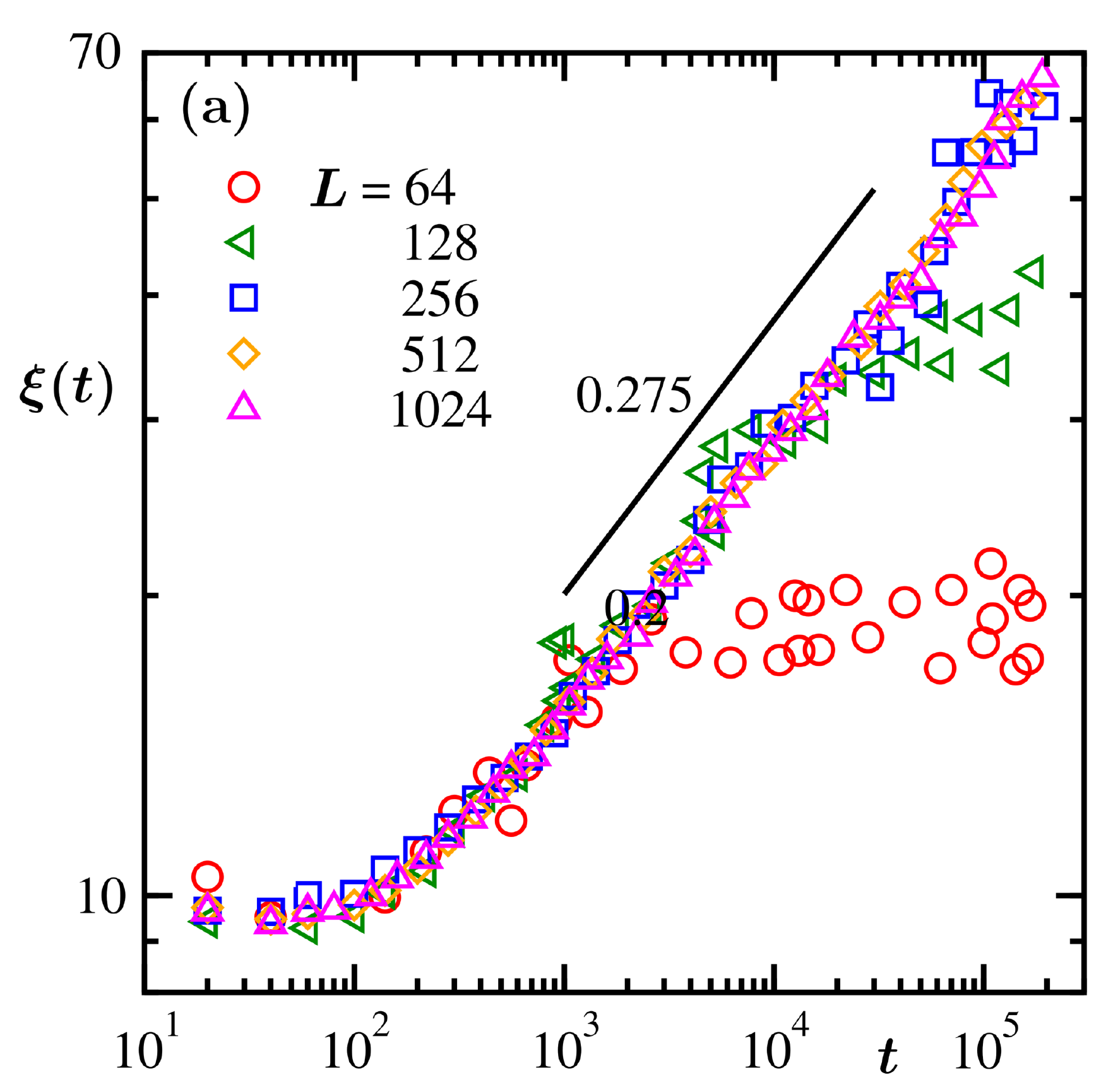}
\includegraphics[width=0.8\linewidth]{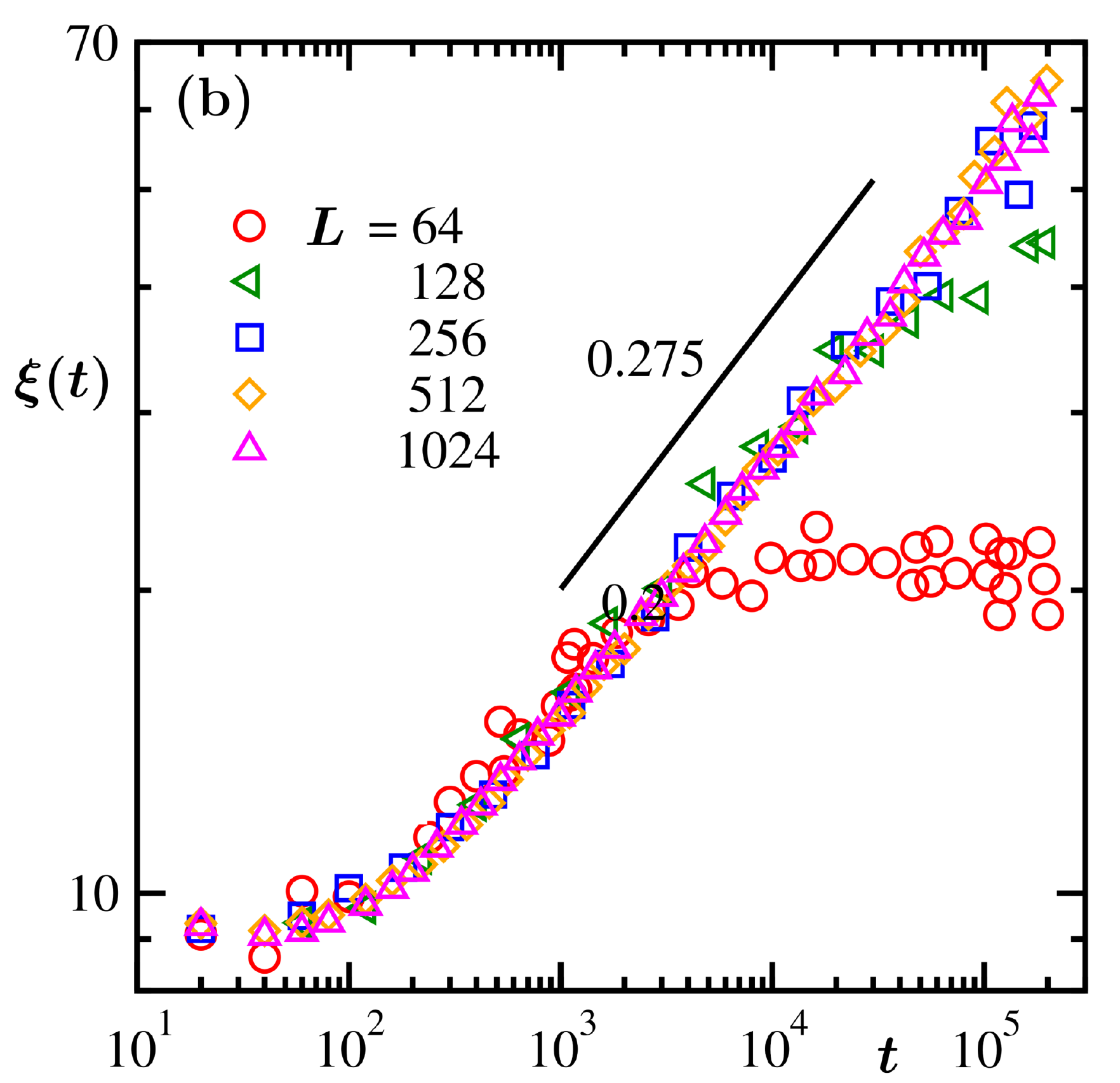}
\includegraphics[width=0.8\linewidth]{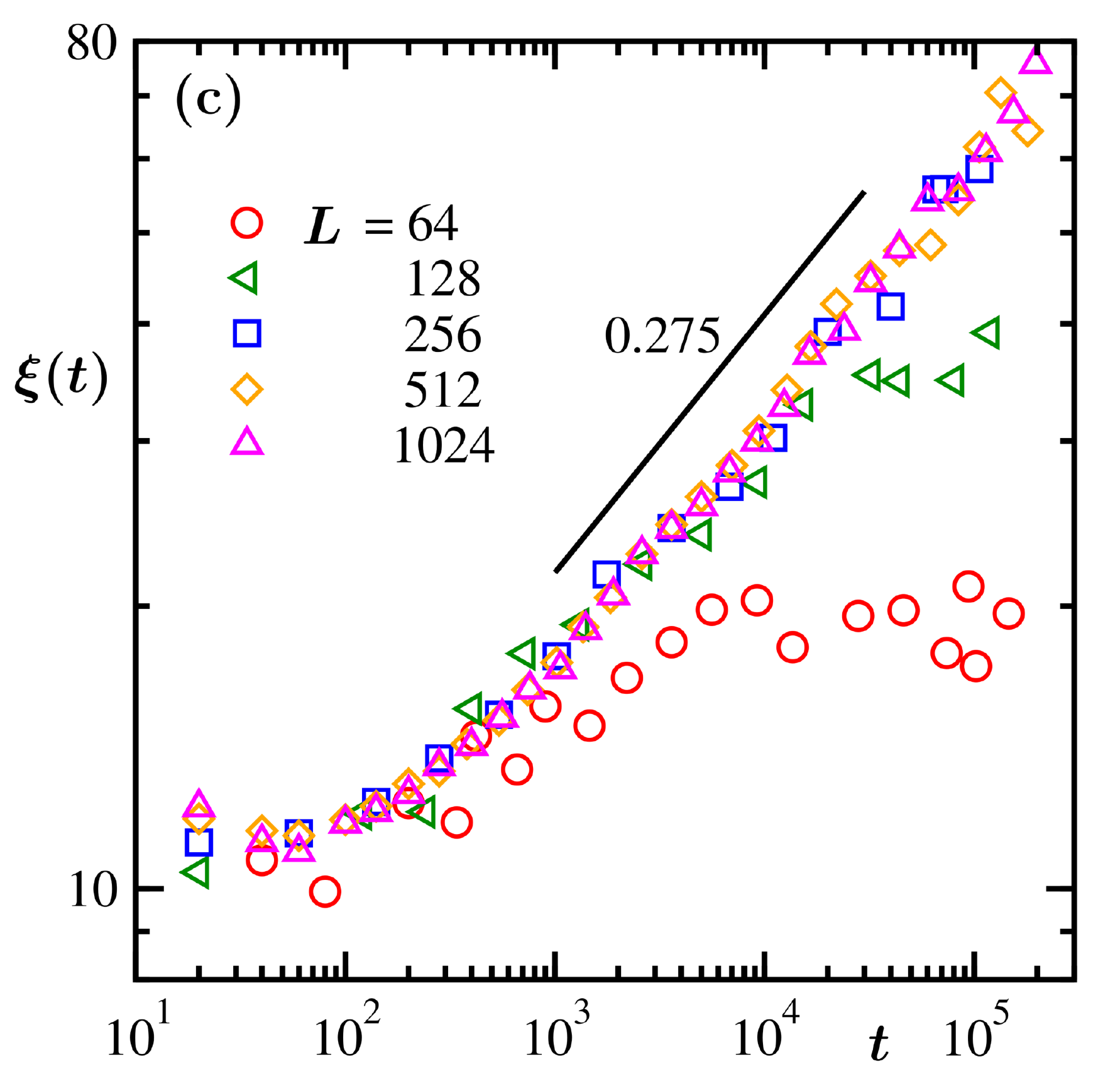}
\caption{Time dependent correlation lengths are plotted for (a) square lattice using method-I, (b) hexagonal lattice using method-I, and  (c) square lattice using method-II. The solid lines are power-laws with mentioned values of the exponent. These results are for quenches of random initial configurations to the critical points.}
\label{fig:corrLength-sq+hx-lattice}
\end{figure}

In Fig.~\ref{fig:SABP} we show snapshots from the evolution of our off-lattice ABP model system following a quench of random initial configurations to various state points. There, locations of particles are marked by dots. The frames under (a) and (b) are for quenches to state points inside the coexistence region (see Fig.~\ref{fig:phdia}). In part (a) the overall density of particles is close to the vapor branch of the coexistence curve. For this case, as expected, we observe formation and growth of disconnected clusters. In part (b) we have included evolution snapshots corresponding to the critical value of $\phi=\phi_{\rm c}$. In this case we observe an essentially bicontinuous structure. The snapshots in part (c) are for quenches to the critical point, and the resulting fractal nature of the morphology can be appreciated. In the following we will only discuss (b) and (c) before moving to results from the lattice models. Note that in the case of (a), gathering meaningful data would require simulations of very large systems over long periods.

First we discuss the case of a quench to the critical point. In this case, due to the fractal nature of the structure the scaling property of Eq. (\ref{eq:scl_cofr}) should be written as
\begin{equation}
C(r,t)\equiv r^{d_f-d}C(r/\xi (t)),
\label{eq:scl_cofr_frac}
\end{equation}
where $d_f$ is the fractal dimension. Recalling that the equilibrium (here steady-state) correlation function in the critical vicinity has the form \cite{Onuki:2002, Fisher_1967}
\begin{equation}
r^{-p} e^{-r/\xi}; ~p=d-2+\eta,
\label{eq:oz_form}
\end{equation}
we have 
\begin{equation}
d_f=2-\eta,
\label{eq:df_and_eta}
\end{equation}
the critical exponent $\eta$ being $1/4$ in space dimension $d=2$ for the Ising class \cite{Fisher_1967}. To check for the scaling property we have thus plotted $r^{0.25} C(r,t)$ as a function of $r/\xi(t)$ in Fig.~\ref{fig:corrfn_offLattice_cr}. Results from several different times have been included and the collapse for $\eta=0.25$ appears good.

The values of $\xi(t)$ obtained via the above discussed scaling analysis are plotted in Fig.~\ref{fig:dLen+xiMax-offLatt-cr}(a), as a function of $t$. Data from different system sizes, as seen on the log-log scale, indicate a power-law growth with the exponent $\simeq 0.275$. This is consistent with $1/z$, with $z=4-\eta$, as expected for the 2D Ising class. In the inset we have shown the instantaneous exponent \cite{Huse:1986, Amar:1988, Suman_Ising:2011}
\begin{equation}
1/z_i=\frac{d \ln \xi(t)}{d\ln t},
\label{eq:insExp_1byzi}
\end{equation}
as a function of $1/\xi(t)$. Asymptotically, a convergence towards $0.275$ can be appreciated. In part (b) of Fig.~\ref{fig:dLen+xiMax-offLatt-cr} we demonstrate that the maximum correlation length scales with the system size, as in the passive case, at the critical point. A more accurate study calls for an exercise where $\xi_{\rm max}$ for different system sizes $L$ are calculated at the finite-size critical points. Now we discuss the case of part (b) in Fig.~\ref{fig:SABP}.

 To check for the self-similar nature of the evolving pattern we calculate $C(r,t)$. Scaling plots of this quantity are presented in Fig.~\ref{fig:scaled-corrfn+sstfrac-offLatt-cr}(a). In this case we aim to validate the scaling form of Eq. (\ref{eq:scl_cofr}). Data from a few different times are shown. There, the distance axis is scaled by the average domain lengths at corresponding times. Clearly, data from different times nicely collapse on top of each other, confirming self-similar growth. A scaling plot for the structure factor is presented in Fig.~\ref{fig:scaled-corrfn+sstfrac-offLatt-cr}(b). There the power-law decay in the large wave vector ($k$) limit validates the Porod law \cite{GPorod1982}. The latter originates from scattering from sharp interfaces. We will discuss the small $k$ power-law behavior later. Note that the presented scaling form for the structure factor $S(k,t)$ is a direct consequence of the fact that this quantity is the Fourier transform of $C(r,t)$.

The average domain lengths are plotted in Fig.~\ref{fig:dLeng+insExp-offLatt-ps}(a) as a function of time. The late time behavior is consistent with a power-law exponent $1/3$. The latter is expected for diffusive domain growth as seen in Lifshitz-Slyozov mechanism \cite{lifshitz:1961} and is realized in Monte Carlo simulations \cite{landau_binder_2005} of Ising model via Kawasaki exchange \cite{kawasaki:1972} kinetics that preserves the system integrated order parameter over time \cite{Bray:2002, landau_binder_2005}. In part (b) of Fig.~\ref{fig:dLeng+insExp-offLatt-ps} we show \cite{Huse:1986, Amar:1988, Suman_Ising:2011}
\begin{equation}
\alpha_i=\frac{d \ln \ell(t)}{d\ln t},
\label{eq:insExp_alphai}
\end{equation}
versus $1/\ell$. Clearly the asymptotic convergence ($\ell=\infty$ limit) is towards a value very close to $1/3$.

In Fig.~\ref{fig:autocorrFn-offLatt-ps} we present the autocorrelation function, $C_{\rm ag} (t,t_w)$, versus the translated time $t-t_w$. Clearly, results from different $t_w$ do not overlap, as expected for evolving systems. The same data sets are plotted versus $t/t_w$ in Fig.~\ref{fig:scaled-autocorrFn-offLatt-ps}. Good overlap is observed. At large values of $t/t_w$ it appears that $C_{\rm ag}$ decays in a power-law manner with an exponent $1$. For an accurate estimate of the exponent in Fig.~\ref{fig:insExp-autocorrFn-offLatt-ps} we show the corresponding instantaneous exponent \cite{DFisher:1988, Jiarul:2015} $-d\ln C_{\rm ag}/d\ln (t/t_w)$ as a function of $t_w/t$. The convergence is towards $1.1$. This implies $\lambda \simeq 3.3$ which is in agreement with the Ising value for conserved order parameter \cite{Jiarul:2015}. In the large $t/t_w$ limit the data set deviates from this scaling behavior due to finite-size effects \cite{Jiarul:2015}, as can be recognized from the direct plot.

For aging dynamics there exist important bounds on the values of $\lambda$. Fisher and Huse (FH) suggested \cite{DFisher:1988}
\begin{equation}
\frac{d}{2}\le \lambda \le d.
\label{eq:lambdaBound}
\end{equation}
The lower bound can be obtained from the well-known Ohta-Jasnow-Kawasaki (OJK) correlation function involving two space points and two times. Recalling that the OJK function \cite{ojk:1982} applies to non-conserved order-parameter dynamics it is expected that the FH lower bound should work for the latter type of dynamics. Later Yeung, Rao and Desai (YRD) \cite{Desai:1996} provided a more general lower bound, viz.,
\begin{equation}
\lambda \ge \frac{d+\beta}{2},
\label{eq:yrdb}
\end{equation}
where $\beta$ is the exponent characterizing the small wave-vector ($k$) power-law behavior \cite{yeung:1988},
\begin{equation}
S(k,t)\sim k^{\beta},
\label{eq:sofk}
\end{equation}
of the equal-time structure factor $S(k,t)$, Fourier transform of $C(r,t)$. Say, for Ising type systems, for standard non-conserved dynamics \cite{ojk:1982, Jiarul:2015} $\beta=0$. Thus, the YRD bound matches with the lower bound of FH. On the other hand for similar models with conserved order-parameter dynamics one should ideally have \cite{yeung:1988} $\beta=4$. 

Note that we are dealing with conserved order-parameter dynamics \cite{Bray:2002, landau_binder_2005} here. In Fig.~\ref{fig:scaled-corrfn+sstfrac-offLatt-cr}(b) we have shown a representative plot of the structure factor, as a function of $k$, on a double-log scale. The small $k$ behavior is consistent with $\beta=3$. In that case we have the YRD bound to be equal to $2.5$, recalling that here $d=2$. Our result in Fig.~\ref{fig:insExp-autocorrFn-offLatt-ps} satisfies the lower bound of YRD. Somewhat smaller value of $\beta$ than $4$ was realized in earlier works also \cite{SAhmad2012}.

Before concluding, we present results from growth in the lattice models. In Fig.~\ref{fig:SLCR} we show evolution snapshots for quenches to the critical points for different lattice models. In Fig.~\ref{fig:corrLength-sq+hx-lattice} we have shown the growth of $\xi$ for these lattice models. The results are consistent with the Ising case. For quenches inside the coexistence regions, patterns obtained from the lattice models differ from the 2D conserved Ising model, and in the late stages the underlying lattice geometry becomes apparent. See Figs.~S1-S2 in Supplementary Material. The average domain length grows faster as well (Fig.~S3). Further investigations, thus, are certainly warranted.

\section{Conclusion}
We have studied kinetics of vapor-liquid phase transition in a model system consisting of Active Brownian particles \cite{Siebert:2018}. Results are presented for structure, growth and aging. Each of these aspects appear to be quite similar to those observed during phase separation in the Ising model with conserved order-parameter dynamics \cite{Bray:2002, landau_binder_2005}. The growth of average domain size follows a power-law behavior with an exponent $\alpha=1/3$, as expected for Lifshitz-Slyzov mechanism \cite{lifshitz:1961}. The aging exponent $\lambda$ appears to have a value $3.3$ that is in quite good agreement with two-dimensional conserved dynamics of Ising model \cite{Jiarul:2015} within $10\%$. The value of $\lambda$ satisfies the Yeung-Rao-Desai bound \cite{Desai:1996}. The structure also matches the Ising behavior.

We have also presented results from a few lattice models \cite{Partridge:2019} for quenches to the critical points. In these cases also the structure and dynamics, like in the case of the continuum model, are similar to those for the conserved Ising model in $d=2$. 

\begin{acknowledgments}
F.D. and P.V. gratefully acknowledge financial support by the Deutsche Forschungsgemeinschaft within priority program SPP 1726 (Grant No. VI 237/5-2) and through Project No. 233630050-TRR 146. SKD acknowledges hospitality during scientific visits in University of Mainz and partial support from JNCASR. Stimulating discussions with Thomas Speck are gratefully acknowledged.
\end{acknowledgments}


\bibliography{bib}%
\end{document}